\documentclass[apjl, onecolumn]{emulateapj}
\def\a{\mathrm{A}}
\def\b{\mathrm{B}}
\def\c{\mathrm{C}}

\newcommand{\secq}{\hspace{0.166667em}}

\newcommand{\out}{\mathrm{out}}

\newcommand{\kbjd}{\mathrm{BJD}-2454833}

\newcommand{\lfill}{\leavevmode%
\leaders\hrule depth-2.1pt height 2.3pt\hfill\kern0pt
}
\usepackage{color}
\usepackage{bm}
\usepackage{aas_macros} 
\usepackage{comment}

\usepackage{float}
\usepackage{amsmath,amssymb}
\usepackage{fancybox}
\usepackage{url}
\usepackage{booktabs}
\usepackage{threeparttable}

\slugcomment{Draft version, June 5}
\shortauthors{Masuda et al.}
\shorttitle{Tertiary Eclipses in KIC 6543674}
\begin{document}
\title{ABSOLUTE DIMENSIONS OF A FLAT HIERARCHICAL TRIPLE SYSTEM KIC 6543674\\ FROM THE \textit{Kepler} PHOTOMETRY}
\author{
Kento \textsc{Masuda}\altaffilmark{1}\altaffilmark{\dag},
Sho \textsc{Uehara}\altaffilmark{2}, and
Hajime \textsc{Kawahara}\altaffilmark{3}
} 
\altaffiltext{1}{
Department of Physics, The University of Tokyo, Tokyo 113-0033, Japan
}
\altaffiltext{2}{
Department of Physics, Tokyo Metropolitan University, 
Tokyo 192-0397, Japan
}
\altaffiltext{3}{
Department of Earth and Planetary Science, The University of Tokyo, Tokyo 113-0033, Japan
}
\email{$\dag$ masuda@utap.phys.s.u-tokyo.ac.jp}
\begin{abstract}
Many of the {\it Kepler} close binaries are suggested to constitute hierarchical triple systems
through their eclipse timing variations (ETVs).
Eclipses by the third body in those systems, if observed, provide precise constraints
on its physical and orbital properties, which are otherwise difficult to obtain.
In this Letter, we analyze such a ``tertiary event" observed only once in the KIC 6543674 system.
The system consists of a short-period ($2.4{\secq}\mathrm{days}$) inner eclipsing binary
and a third body on a wide ($1100{\secq}\mathrm{days}$) and eccentric ($e\simeq0.6$) orbit.
Analysis of three tertiary eclipses around a single inferior conjunction of the third body
yields the mutual inclination between the inner and outer binary planes
to be $3\fdg3\pm0\fdg6$, indicating an extremely flat geometry.
Furthermore, combining the timings and shapes of the tertiary eclipses with the phase curve
and ETVs of the inner binary,
we determine the mass and radius ratios of all three bodies in the system
using the {\it Kepler} photometry alone.
With the primary mass and temperature from the {\it Kepler} Input Catalog,
the absolute masses, radii, and effective temperatures of the three stars are obtained as follows:
$M_\a=1.2\pm0.3{\secq}M_\odot$, $R_\a=1.8\pm0.1{\secq}R_\odot$,
$M_\b=1.1_{-0.2}^{+0.3}{\secq}M_\odot$, $R_\b=1.4\pm0.1{\secq}R_\odot$,
$M_\c=0.50_{-0.08}^{+0.07}{\secq}M_\odot$, $R_\c=0.50\pm0.04{\secq}R_\odot$,
$T_\a\simeq{T_\b}\simeq6100{\secq}\mathrm{K}$, and $T_\c<5000{\secq}\mathrm{K}$.
Implication for the formation scenario of close binaries is briefly discussed.
\end{abstract}

\section{Introduction}\label{sec:intro}
Among over $2000$ eclipsing binaries discovered in the {\it Kepler} mission 
\citep{2011AJ....141...83P, 2011AJ....142..160S},
more than $200$ are suggested to host tertiary (third body) companions
through their eclipse timing variations
\citep[ETVs;][]{2014AJ....147...45C}. 
Many of them are hierarchical triples
consisting of a short-period binary and an outer third body on a wide orbit.
The hierarchy is often attributed to the perturbation from the third body,
as in the well-known KCTF (Kozai cycles with tidal friction) scenario \citep{1962AJ.....67..591K, 1998MNRAS.300..292K,  2001ApJ...562.1012E}.
Indeed, recent ETV analyses \citep{2013ApJ...768...33R, 2015MNRAS.448..946B}
have revealed many hierarchical triples with misaligned tertiary orbits, 
whose mutual inclinations exhibit suggestive peaks around $\sim 40^\circ$
in agreement with the KCTF prediction \citep{2007ApJ...669.1298F}.

On the other hand, at least $10$ or more hierarchical triples seem to have well-aligned orbits,
as suggested by eclipses due to tertiary companions
\citep[][figure 7]{2011Sci...331..562C, 2015arXiv150307295O}.
Three-dimensional geometry and absolute dimensions of those systems
are also of interest because their hierarchy may argue for the mechanism of orbital shrinkage
that do not require high mutual inclinations
between the inner and outer binary planes \citep[e.g.,][]{2015ApJ...805...75P}.

\begin{figure*}[htbp]
	\centering
	\rotatebox{90}{
		\includegraphics[width=9.5cm, clip]{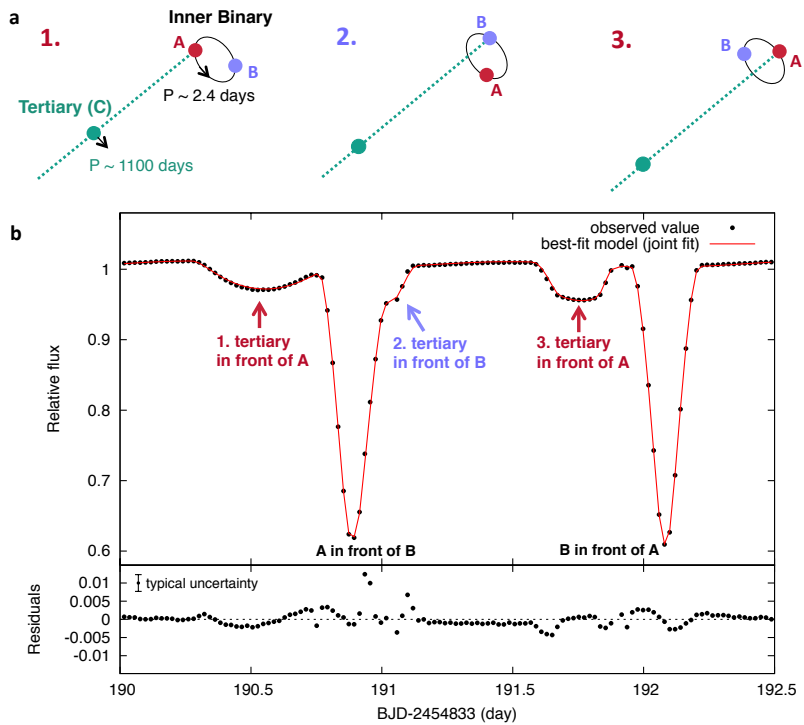}
	}
	\caption{Tertiary event observed in the KIC 6543674 system and its interpretation.
	(a) Schematic illustration of the system configuration during the event.
	(b) Fit to the {\it Kepler} light curve around the tertiary eclipses (see Section \ref{sec:tertiary}).
	({\it Top}) Black circles are the observed fluxes and red solid line denotes our best-fit model.
	({\it Bottom}) Residuals of our fit. 
	Typical uncertainty estimated from our analysis ($\simeq\sigma_\mathrm{LC,tertiary}$)
	is shown at the upper left.}\label{fig:tertiary}
\end{figure*}

In this Letter, we focus on a tertiary event observed only once in the KIC 6543674 system,
which involves three tertiary eclipses around a single inferior conjunction of the third body (Figure \ref{fig:tertiary}).
Although this event has already been reported 
\citep{2011AJ....142..160S, 2013AAS...22114209T, 2014AJ....147...45C},
the information obtained from its detailed modeling has not yet been clarified.
Below we will show that the tertiary event plays two crucial roles in determining 
the system configuration.
First, it constrains the mutual inclination between the inner and outer binary orbits very precisely,
in a similar way to the ``planet--planet eclipse" known in the {\it Kepler} multi-transiting planetary system(s)
\citep[][]{2012ApJ...759L..36H, 2013ApJ...778..185M, 2014ApJ...783...53M}.
Secondly, and less trivially, it 
fixes the mass ratio of the inner binary and velocity of the third body even without spectroscopy.

The present Letter reports precise geometry and absolute dimensions of the KIC 6543674 system.
We combine the above information from the tertiary event with the complementary constraints from ETVs and eclipses of the inner binary.
To obtain a consistent solution,
we fit the three components simultaneously using a Markov Chain Monte Carlo (MCMC) method.
Section \ref{sec:etv_phase} presents individual analyses of the ETVs and eclipse curves of the inner binary.
We then model the two components jointly with the tertiary eclipses in Section \ref{sec:tertiary}
to determine the parameters of the whole system.
Section \ref{sec:discussion} discusses the implication of the resulting system architecture 
and the prospects for the follow-up observations to better understand this valuable system.\clearpage

\begin{deluxetable*}{lcccc}
	\tabletypesize{\small}
	\tablewidth{0pt}
	\tablecolumns{5}
	\tablecaption{System parameters from the {\it Kepler} light curves.}
	\tablehead{
	Parameter & \colhead{ETVs}  & \colhead{phase curve} & \colhead{ETVs + phase + tertiary} 
	& \colhead{ETVs + phase + tertiary}\\
	&&&
	& \colhead{(with the prior on $M_\a$ from KIC)}
	}
	\startdata
	\multicolumn{5}{c}{({\it Inner binary}) \lfill}\\
	$t_{0, \rm in}$ ($\kbjd$)		& $132.3070\pm0.0002$	& $\cdots$	& $132.3071\pm0.0001$ & $132.30704\pm0.00009$\\
	$t_{0, \rm in}^{\rm phase}$ ($\kbjd$)	& $\cdots$ & $132.30372\pm0.00004$ & $132.30372\pm0.00003$ & $132.30372_{-0.00003}^{+0.00002}$\\
	$P_{\rm in}$ (day)		& $2.3910305\pm0.0000003$ & $2.3910305$ (fixed) & $2.3910305\pm0.0000003$ & $2.3910305\pm0.0000002$\\
	$a_{\rm in}/R_\a$		& $\cdots$			& $5.49\pm0.02$		& $5.494_{-0.006}^{+0.007}$ & $5.494_{-0.007}^{+0.006}$\\
	$\cos i_{\rm in}$				& $\cdots$	& $0.021\pm0.002$		& $0.022\pm0.002$ & $0.022\pm0.002$\\
	$e_{\rm in} \cos \omega_{\rm in}$	& $\cdots$	& $(0.2\pm3.3)\times10^{-5}$ & $0$ (fixed) & $0$ (fixed)\\
	$e_{\rm in} \sin \omega_{\rm in}$	& $\cdots$	& $-0.0005_{-0.0020}^{+0.0021}$	      & $0$ (fixed) & $0$ (fixed)\\
	$R_\b/R_\a$					& $\cdots$	& $0.781\pm0.004$			& $0.781\pm0.002$ & $0.781\pm0.002$\\
	$M_\b/M_\a$					& $\cdots$	& $\cdots$				& $0.93\pm0.02$     & $0.93\pm0.02$\\
	$C_{\rm phase}$				& $\cdots$	& $1.00259\pm0.00002$		& $1.00259\pm0.00002$ & $1.00259\pm0.00002$\\
	$T_\b/T_\a$					& $\cdots$	& $1.012\pm0.002$		& $1.0107\pm0.0004$ & $1.0107\pm0.0004$\\
	$u_\a$						& $\cdots$	& $0.45\pm0.04$			& $0.434\pm0.009$ & $0.434\pm0.009$\\
	$u_\b$						& $\cdots$	& $0.46\pm0.03$			& $0.47\pm0.02$     & $0.46\pm0.02$\\
	$A_0$						& $\cdots$	& $0.041\pm0.007$			& $0.037\pm0.006$ & $0.037\pm0.006$\\
	$A_{1\rm c}$					& $\cdots$	& $0.00034\pm0.00005$		& $0.00035\pm0.00005$ & $0.00035\pm0.00005$\\
	$A_{1\rm s}$					& $\cdots$	& $0.00096\pm0.00004$		& $0.00096\pm0.00004$ & $0.00096\pm0.00004$\\
	$A_{2\rm c}$					& $\cdots$	& $-0.00720\pm0.00007$		& $-0.00716\pm0.00006$ & $-0.00716\pm0.00006$\\
	$R_\a$ ($R_\odot$)				& $\cdots$	& $\cdots$& $2.1_{-0.8}^{+3.2}$\tablenotemark{\dag}   & $1.8\pm0.1$\tablenotemark{\dag}\\
	$R_\b$ ($R_\odot$)				& $\cdots$	& $\cdots$& $1.6_{-0.7}^{+2.5}$\tablenotemark{\dag}   & $1.4\pm0.1$\tablenotemark{\dag}\\
	$M_\a$ ($M_\odot$)				& $\cdots$	& $\cdots$	& $1.8_{-1.4}^{+27.5}$\tablenotemark{\dag} & $1.2\pm0.3$\\
	$M_\b$ ($M_\odot$)				& $\cdots$	& $\cdots$	& $1.7_{-1.3}^{+25.5}$\tablenotemark{\dag} & $1.1_{-0.2}^{+0.3}$\tablenotemark{\dag} \\
	\multicolumn{5}{c}{({\it Third body}) \lfill}\\
	$t_{0, \out}$ ($\kbjd$)		& $199\pm10$		& $\cdots$	& $191.246\pm0.003$	& $191.246\pm0.003$\\
	$P_\out$ (day)			& $1086_{-7}^{+8}$	& $\cdots$	& $1090\pm6$			& $1090\pm5$\\
	$e_\out \cos \omega_\out$	& $0.13\pm0.05$	& $\cdots$	& $0.16\pm0.03$		& $0.16\pm0.03$\\
	$e_\out \sin \omega_\out$		& $0.58\pm0.03$	& $\cdots$	& $0.58\pm0.02$		& $0.572\pm0.008$\\
	$a_\out /R_\a$			& $\cdots$		& $\cdots$	& $345_{-13}^{+15}$		& $348\pm2$\tablenotemark{\dag}\\
	$\cos i_\out$				& $\cdots$		& $\cdots$	& $0.0030\pm0.0003$	& $0.0029_{-0.0002}^{+0.0001}$\\
	$\Delta\Omega$ (deg)	& $\cdots$		& $\cdots$	& $3.2\pm0.6$		& $3.1\pm0.6$\\
	$A_{\rm ETV}$	(s)		& $264\pm6$		& $\cdots$	& $266\pm5$		& $265\pm5$\tablenotemark{\dag}\\
	$C_{\rm tertiary}$		& $\cdots$		& $\cdots$	& $1.0070\pm0.0003$	& $1.0070\pm0.0003$\\
	$\gamma_{\rm tertiary}$ ($\mathrm{day}^{-1}$) & $\cdots$	& $\cdots$	& $0.00004\pm0.00021$ & $0.00005_{-0.00022}^{+0.00021}$\\
	$R_\c/R_\a$			& $\cdots$		& $\cdots$	& $0.277\pm0.003$		& $0.277\pm0.003$\\
	$M_\c/M_\a$			& $\cdots$		& $\cdots$	& $0.4_{-0.2}^{+0.3}$\tablenotemark{\dag} & $0.43_{-0.03}^{+0.04}$\\
	$T_\c/T_\a$			& $\cdots$		& $\cdots$& $0.84_{-0.04}^{+0.03}$\tablenotemark{\dag} & $0.84_{-0.04}^{+0.03}$\tablenotemark{\dag}\\
	$R_\c$ ($R_\odot$)		& $\cdots$		& $\cdots$& $0.6_{-0.2}^{+0.9}$\tablenotemark{\dag} & $0.50\pm0.04$\tablenotemark{\dag}\\	
	$M_\c$ ($M_\odot$)		& $\cdots$		& $\cdots$& $0.7_{-0.4}^{+3.3}$\tablenotemark{\dag} & $0.50_{-0.08}^{+0.07}$\tablenotemark{\dag}\\	
	mutual inclination (deg)	& $\cdots$ & $\cdots$ & $3.3\pm0.6$\tablenotemark{\dag} & $3.3_{-0.6}^{+0.5}$\tablenotemark{\dag}\\	
	\multicolumn{5}{c}{({\it Jitters}) \lfill}\\
	$\sigma_{\rm ETV}$ ($\mathrm{s}$)	 & $56\pm3$	& $\cdots$ & $56\pm3$	& $56\pm3$\\ 
	$\sigma_{\rm LC, phase}$	 & $\cdots$	& $0.00048\pm0.00001$		& $0.00049\pm0.00001$		& $0.00049\pm0.00001$\\
	$\sigma_{\rm LC, tertiary}$ & $\cdots$	& $\cdots$				& $0.0023\pm0.0002$		& $0.0023_{-0.0001}^{+0.0002}$
	\enddata
	\tablecomments{The quoted values and uncertainties are the median and $68.3\%$ credible interval
	of the marginalized posteriors.
	Values marked with daggers are derived from the posteriors of other fitted parameters.}\label{tab:results}
\end{deluxetable*}

\section{Constraints from ETVs and Phase Curve of the Inner Binary}\label{sec:etv_phase}
The KIC 6543674 system consists of the inner eclipsing binary with the orbital period of $P_\mathrm{in}\simeq2.39\secq\mathrm{days}$
and the ``outer" eccentric binary (third body moving around the center of mass of the inner binary)
with $P_\out\simeq1100\secq\mathrm{days}$.
Here we present individual MCMC analyses of the phase curve and ETVs of the inner binary,
which allow us to constrain the orbital geometries of the inner and outer binaries, respectively.
Since $P_\mathrm{in}/P_\out$ is small, 
both inner and outer binary orbits are approximately Keplerian.
We adopt the approximation throughout the paper and define
all the orbital elements in Jacobi coordinates (with subscripts ``in" and ``out"), which are in this case constant over time.

\subsection{ETV Analysis}\label{ssec:etv}
The inner binary exhibit ETVs, 
which were used to infer the existence of the third body \citep{2014AJ....147...45C}.
They are caused by the finite light-travel time (R{\o}mer delay) and the variation in the line-of-sight distance 
due to the outer binary motion. 
Under our assumption, the $i$th eclipse time of the inner binary $t_i$ can be modeled as \citep{2013ApJ...768...33R}\footnote{The sign is opposite to their equation (6) because we take $+z$-axis in the observer's direction.}\begin{align}t^\mathrm{model}_i=t_{0,\mathrm{in}}+P_\mathrm{in}i+A_\mathrm{ETV}\left\{\sqrt{1-e_\out^2}\sin E_\out(t_i)\cos\omega_\out+[\cos E_\out(t_i)-e_\out]\sin\omega_\out\right\}.\label{eq:eclipse_times}\end{align}Here, $t_{0,\mathrm{in}}$ is the eclipse epoch (time of inferior conjunction) of the inner binary,
and $e_\out$, $\omega_\out$, and $E_\out$ are the eccentricity, argument of pericenter, and eccentric anomaly
of the third body.
The amplitude of ETVs, $A_\mathrm{ETV}$, is given by the projected semi-major axis 
of the outer binary $a_\out\sin i_\out$ divided by the speed of light $c$:\begin{align}A_\mathrm{ETV}=\frac{(GM_\a)^{1/3}}{c(2\pi)^{2/3}}{\secq}\frac{(M_\c/M_\a)\sin i_\out}{(1+M_\b/M_\a+M_\c/M_\a)^{2/3}}{\secq}P_\out^{2/3},\label{eq:Aetv}\end{align}
where $M$ denotes the stellar mass, with the subscripts $A$, $B$, and $C$
specifying the primary, secondary, and tertiary stars, respectively.
In such a hierarchical system as KIC 6543674, dynamical effects that change $P_{\rm in}$
are sufficiently smaller than the above effect and so are neglected \citep{2013ApJ...768...33R}.

\begin{figure*}[htbp]
	\centering
	\rotatebox{90}{
	\includegraphics[width=6.4cm, clip]{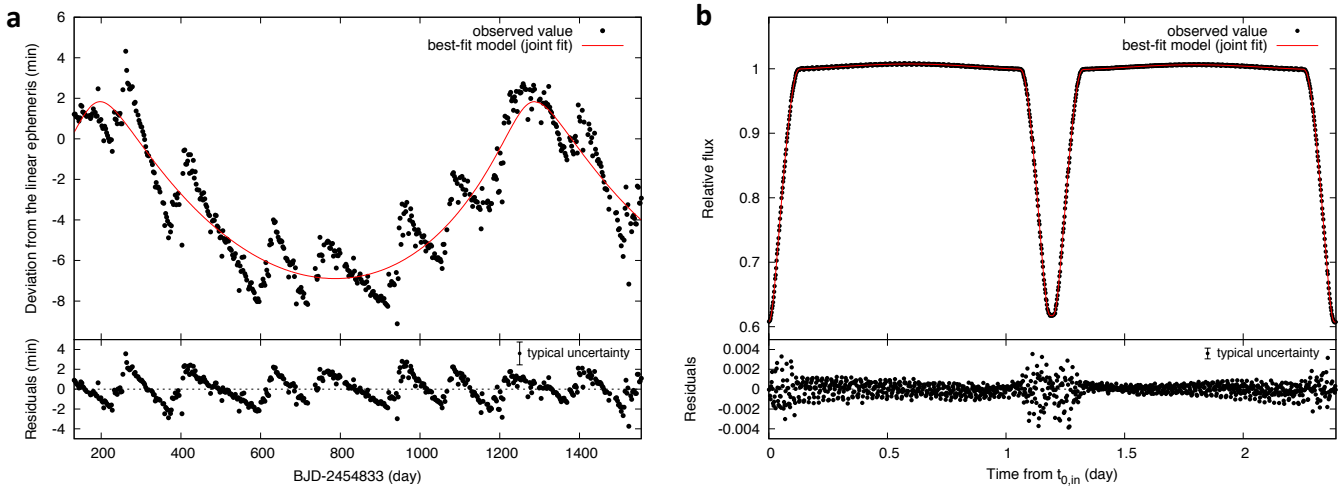}
	}
	\caption{(a) Fit to the eclipse times. ({\it Top}) Black circles are the observed eclipse times and red solid line denotes our best-fit model.
	Only the deviations from the linear ephemeris are shown for clarity.
	({\it Bottom}) Residuals of our fit. Typical (jitter-included) uncertainty is shown at the upper right.
	(b) Fit to the folded phase curve. 
	({\it Top}) Black circles are the observed fluxes and red solid line denotes our best-fit model.
	({\it Bottom}) Same as panel (a).}\label{fig:etv_phase}
	\vspace{0.6cm}
\end{figure*}

We use Equation (\ref{eq:eclipse_times}) 
to model the primary eclipse times $t_i^\mathrm{obs}$ in table 1 of \citet{2014AJ....147...45C}
obtained by fitting the light curve over the entire phase (flagged as ``entire").
The observed ETVs also exhibit short-term modulations 
(see Figure \ref{fig:etv_phase}a), which can be explained by star spots
if the stellar rotation is nearly (but not exactly) synchronized with the inner binary motion
\citep[see, e.g., figure 3 of][]{2015arXiv150307295O}.
Instead of modeling them, 
we include an additional scatter $\sigma_\mathrm{ETV}$ to the formal eclipse-time error $\sigma_i$ in quadrature
to define the following likelihood for the ETV fit:
\begin{equation}\mathcal{L}_\mathrm{ETV}=\prod_{i}\frac{1}{\sqrt{2\pi(\sigma_i^2+\sigma_\mathrm{ETV}^2)}}\exp\left[\frac{(t_i^\mathrm{obs}-t_i^\mathrm{model})^2}{2 (\sigma_i^2+\sigma_\mathrm{ETV}^2)}\right].\label{eq:Letv}\end{equation}
This likelihood is used to perform an MCMC sampling \citep[{\tt emcee} by][]{2013PASP..125..306F}
of the posteriors of 
the parameters in the second column of Table \ref{tab:results}.
The best-fit model is compared with the observed values in Figure \ref{fig:etv_phase}a.

\subsection{Phase-Curve Analysis}\label{ssec:phasecurve}
The linear ephemeris of the inner binary ($t_{0,\mathrm{in}}$ and $P_\mathrm{in}$) obtained in Section \ref{ssec:etv}
is used to phase-fold the light curve taken from the {\it Kepler} eclipsing binary catalog,\footnote{\url{http://keplerebs.villanova.edu}}
whose instrumental trend has been removed (``flattened") using polynomials \citep{2014AJ....147...45C}.
Since $A_\mathrm{ETV}$ is shorter than the data cadence ($29.4{\secq}\mathrm{minutes}$),
we do not correct for ETVs here and in the following light-curve fitting (Section \ref{sec:tertiary}).
The folded fluxes are averaged into three minute bins, 
and the flux value and error in each bin are estimated as the median and $1.4826$ times median absolute deviation divided by 
the square root of the number of points in the bin.

We model the flux over the entire phase as
\begin{equation}f_\mathrm{phase}(t)=\frac{C_\mathrm{phase}}{1+F_\b/F_\a+A_0}\left[f_\a(t)+\frac{F_\b}{F_\a}{\secq}f_\b(t)+A_0+A_{1\mathrm{c}}\cos\phi+A_{1\mathrm{s}}\sin\phi+A_{2\mathrm{c}}\cos2\phi\right].\label{eq:fphase}\end{equation}Here, 
$f_{\a,\b}(t)$ is the normalized stellar flux
computed with the analytic eclipse model by \citet{2002ApJ...580L.171M} for the linear limb darkening law.
They are determined from the orbital ephemeris,
scaled semi-major axis $a_\mathrm{in}/R_\a$,
cosine of the orbital inclination $\cos i_\mathrm{in}$,
radius ratio $R_\b/R_\a$,
and linear limb-darkening coefficients $u_\a$ and $u_\b$.
The flux ratio, $F_\b/F_\a$, is computed by $(R_\b/R_\a)^2(T_\b/T_\a)^4$,
where $T$ is the stellar effective temperature in the {\it Kepler} band.
The constants $A_0$, $A_{1\mathrm{c}}$, $A_{1\mathrm{s}}$, and $A_{2\mathrm{c}}$ 
are the phenomenological parameters to describe the phase-curve modulation,
and $\phi=2\pi(t-t_{0,\mathrm{in}})/P_\mathrm{in}$ is the orbital phase.\footnote{
Since ETVs we neglected may shift the center of the phase curve,
we allow $t_{0,\mathrm{in}}$ used for the phase-curve fitting (denoted as
$t_{0,\mathrm{in}}^\mathrm{phase}$) to be different from $t_{0,\mathrm{in}}$ in Equation (\ref{eq:eclipse_times}).
The resulting difference ($\left|t_{0,\mathrm{in}}^\mathrm{phase}-t_{0,\mathrm{in}}\right|\simeq{\secq}5{\secq}\mathrm{minutes}$) 
is actually comparable to $A_\mathrm{ETV}$ and consistent with the ETV origin.}
These amplitudes, in principle, can be related to the masses of the two bodies
with the physical model of ellipsoidal variation and Doppler beaming 
\citep{1993ApJ...419..344M, 2003ApJ...588L.117L}.
We do not use them for the mass estimates, however, 
because our quarter-by-quarter analysis reveals 
the temporal variation in the shape of the phase curve. 
This variation is also consistent with the star-spot modulation
nearly synchronized with the orbital motion.
Finally, $C_\mathrm{phase}$ is the overall normalization.
In fitting the observed data, $f_\mathrm{phase}(t)$ is averaged over $30$ minutes around each time
to take into account the long-cadence sampling.
The light-travel time effect is neglected in computing $f_\mathrm{phase}(t)$ 
because it is shorter than the data cadence.

As in Section \ref{ssec:etv}, we use an MCMC algorithm to fit the phase-folded light curve  
for the above parameters.
We again include the ``jitter" term $\sigma_\mathrm{LC,phase}$ in the likelihood $\mathcal{L}_\mathrm{phase}$
defined in the same way as in Equation (\ref{eq:Letv}).
The resulting constraints are in the third column of Table \ref{tab:results}, and the best-fit light curve
is shown in Figure \ref{fig:etv_phase}b.
We also try floating $e_\mathrm{in}$ and $\omega_\mathrm{in}$,
only to find that the inner orbit is very close to circular. Hence we fix $e_\mathrm{in}=0$ in the following analyses.

The residuals in the bottom panel of Figure \ref{fig:etv_phase}b exhibit an out-of-eclipse warp and a larger in-eclipse scatter
\citep[similar to the one in][]{2012ApJ...761..157B}.
The former does not affect our analysis significantly because we do not extract any physical information from the out-of-eclipse modulation.
On the other hand, the latter points to systematics that affect the shape of eclipses and thus may bias the resulting system parameters.
While it may be due to the spot occultation, 
ETVs we neglected could also affect the eclipse signal by a similar amount
($A_\mathrm{ETV}/(\mathrm{ingress{\secq}duration})\sim\mathcal{O}(1\%)$).
Although unlikely to explain the random scatter, 
we also note that the \citet{2002ApJ...580L.171M} model is exact only for spherical stars
and so neglects the tidal distortion of $\mathcal{O}(1\%)$ suggested by the value of $A_{2\mathrm{c}}$.
In any case, the results of the following analyses could suffer from that level of systematics,
though the main conclusions remain unchanged.
\section{Geometry and Absolute Dimensions from the Tertiary Event}\label{sec:tertiary}
In this section, we analyze the light curve of the tertiary event 
jointly with the two components in the previous section.
The outer binary motion of the third body is converted to the motions relative to the primary and secondary,
which are used to compute their normalized fluxes including the tertiary eclipses,
$f_{\a,\mathrm{tertiary}}(t)$ and $f_{\b,\mathrm{tertiary}}(t)$, 
with the \citet{2002ApJ...580L.171M} model.
This requires
$a_\out/R_\a$, $\cos i_\out$, $R_\c/R_\a$,
$\Delta\Omega$ (difference in the longitudes of ascending node between inner and outer orbits)
and $M_\b/M_\a$ in addition to the parameters in Section \ref{sec:etv_phase}.
They are incorporated in the model flux during the tertiary event as
\begin{align}f_\mathrm{tertiary}(t)=\frac{C_\mathrm{tertiary}+\gamma_\mathrm{tertiary}(t-t_*)}{1+F_\b/F_\a+A_0}\left[f_{\a,\mathrm{tertiary}}(t)+\frac{F_\b}{F_\a}{\secq}f_{\b,\mathrm{tertiary}}(t)+A_0+A_{1\mathrm{c}}\cos\phi+A_{1\mathrm{s}}\sin\phi+A_{2\mathrm{c}}\cos2\phi\right],\end{align}
where
$C_\mathrm{tertiary}$ is the normalization,
$\gamma_\mathrm{tertiary}$ models the residual instrumental trend  
around the tertiary event, and we choose $t_*(\kbjd)=191.25$.
The model likelihood for the tertiary-event light curve $\mathcal{L}_\mathrm{tertiary}$ is defined
in the same way as in $\mathcal{L}_\mathrm{phase}$,
again including an additional jitter $\sigma_\mathrm{LC,tertiary}$.
We first seek for the solution that maximizes $\mathcal{L}_\mathrm{tertiary}$ 
with $\sigma_\mathrm{LC,tertiary}=0$ for various $t_{0, \out}$ 
using the Levenberg-Marquardt method \citep{2009ASPC..411..251M}.
Here the above seven new parameters are fitted, while the others are floated within
the $3\sigma$ boundaries from the ETVs and phase curve (Table \ref{tab:results}).
We then perform an MCMC run from the solution,
fitting all the model parameters simultaneously with the joint likelihood 
$\mathcal{L}\propto\mathcal{L}_\mathrm{ETV}\cdot\mathcal{L}_\mathrm{phase}\cdot\mathcal{L}_\mathrm{tertiary}$.
The resulting constraints are summarized in the fourth column of Table \ref{tab:results}
along with other derived parameters.
As shown in Figure \ref{fig:tertiary}, our model well reproduces the observed tertiary eclipses.
In the following subsections, we discuss the information newly derived from the tertiary eclipses.  

\subsection{Mutual Inclination}
Tertiary eclipses on both of the inner two stars suggest a well-alignment between inner and outer binary planes.
This naive expectation is quantified by our modeling.
We obtain $i_\out=89\fdg83\pm0\fdg02$
and 
$\Delta\Omega=3\fdg2\pm0\fdg6$ (see Figure \ref{fig:diagram}b) 
as the line-of-sight and sky-plane inclinations of the tertiary orbit.
Combined with $i_\mathrm{in}=88\fdg7\pm0\fdg1$, these results indicate an extremely flat orbital configuration,
with the $3\sigma$ upper limit on the mutual inclination being $5^\circ$.

\subsection{Relative Dimensions}
Another role of the tertiary event is to determine the mass ratio $M_\b/M_\a$ and
the tertiary-to-primary velocity ratio $V_\c/V_\a$ during the event,
where $V$ is the orbital velocity relative to the center of mass of the inner binary.
The constraints are invaluable because they allow us to determine the mass ratios
of all three bodies.
It is even possible, in principle, to combine them with the ETV amplitude 
to fix the absolute dimensions of the whole system from photometry alone.

\begin{figure*}[htbp]
	\centering
	\rotatebox{90}{
		\includegraphics[width=8cm, clip]{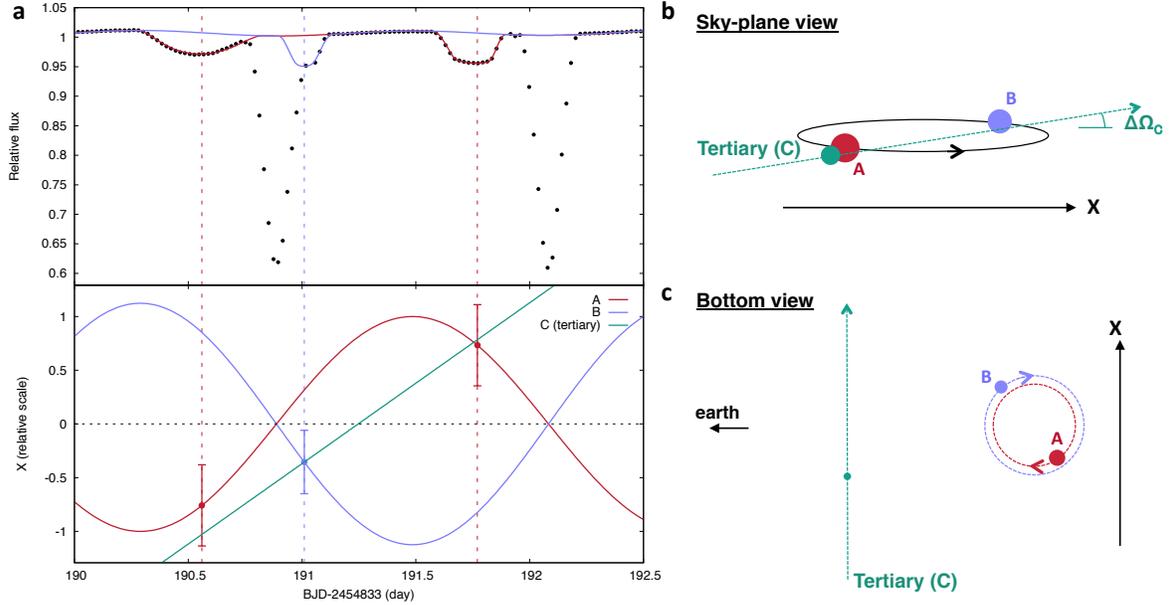}
	}
	\caption{(a) Relationship between the timings of three tertiary eclipses and motions of three stars.
	({\it Top}) The black dots denote the detrended {\it Kepler} light curve.
	The red and blue lines are the best-fit tertiary eclipse models for stars A and B, respectively.
	The vertical dashed lines denote the rough central times of the tertiary eclipses.
	({\it Bottom}) One-dimensional motion of the three stars
	(primary: red, secondary: blue, tertiary: green) with respect to the center of mass of the inner binary.
	The $X$-axis is defined to coincide with the line of nodes of the inner binary, with its positive direction
	shown in panels (b) and (c).
	The amplitude of the primary motion is normalized to unity,
	while that of the secondary depends on $M_\b/M_\a$
	(notice that only the relative scale affects the light curve).
	The vertical bars denote the normalized radii of stars A (red) and B (blue).
	(b) Sky-plane view and (c) bottom view of the system.
	Definitions of $\Delta\Omega$ and $X$-axis are shown schematically.}\label{fig:diagram}
\end{figure*}

The two quantities, $M_\b/M_\a$ and $V_\c/V_\a$,
are closely related to the timings and durations of the three tertiary eclipses.
The bottom panel of Figure \ref{fig:diagram}a shows the 
approximately one-dimensional motion of the inner binary in the sky plane
with respect to its center of mass (red and blue sinusoidal lines).
Here the motion of the third body (green line) is represented by an almost straight line 
owing to its long orbital period.
For $\Delta\Omega\simeq0^\circ$,
eclipses occur at the intersections of the two lines in this diagram.
Thus, the green line should cross either of the red or blue sinusoids 
at the times of three tertiary eclipses (vertical dashed lines), 
roughly within the primary/secondary radii (vertical error bars).
The condition essentially fixes the amplitude of the blue sinusoid and the slope of the green line,
which correspond to $M_\a/M_\b$ and $V_\c/V_\a$, respectively.
The ratio $V_\c/V_\a$ is further constrained by the relative durations of the first and third tertiary eclipses,
where the relative velocities between the two stars are $V_\a-V_\c$ and $V_\a+V_\c$, respectively.

These ratios yield the relative mass of the third body as well.
Using $P_\mathrm{in}$, $a_\mathrm{in}/R_\a$, $t_{0, \out}$, $P_\out$, $e_\out$, and $\omega_\out$ we already derived,
$V_\c/V_\a$ is converted to $a_\out/R_\a$.
Since this $a_\out$ should satisfy Kepler's third law, we obtain
\begin{equation}\left(\frac{a_\out/R_\a}{a_\mathrm{in}/R_\a}\right)^3\left(\frac{P_\mathrm{in}}{P_\out}\right)^2=1+\frac{M_\c/M_\a}{1+M_\b/M_\a},\label{eq:ac/ra}\end{equation}
which can be solved for $M_\c/M_\a$ as
\begin{equation}\frac{M_\c}{M_\a}=\left[\left(\frac{a_\out/R_\a}{a_\mathrm{in}/R_\a}\right)^3\left(\frac{P_\mathrm{in}}{P_\out}\right)^2-1\right]\left(1+\frac{M_\b}{M_\a}\right).\label{eq:mc/ma}\end{equation}
The mass ratios derived in this way are listed in Table \ref{tab:results}.
These values indicate that the system is dynamically stable, according to the criterion by \citet{2001MNRAS.321..398M}.

\begin{figure*}[htbp]
	\centering
	\rotatebox{90}{
		\includegraphics[width=13cm, clip]{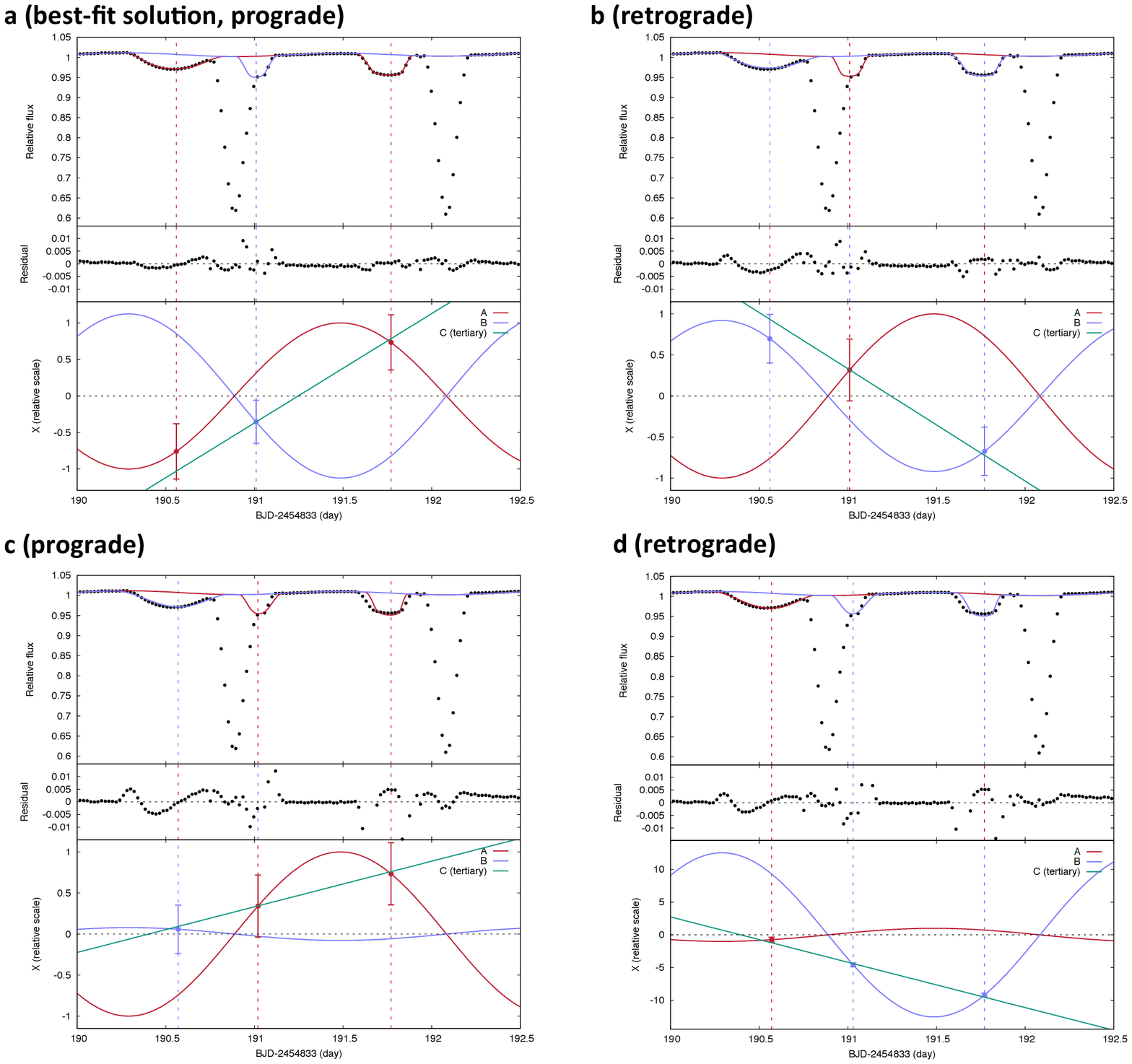}
	}
	\caption{Comparison between the best-fit solution (panel a)
	and other solutions allowed from the timings of the three eclipses alone (panels b, c, and d).
	The meaning of each panel is basically the same as Figure \ref{fig:diagram}a,
	but this time the residuals for each solution is shown in the middle using the same scales.}\label{fig:diagram_others}
\end{figure*}

In fact, the timings of the three eclipses alone allow for other configurations,
though they do not fit the eclipse shapes well and hence are rejected (Figure \ref{fig:diagram_others}).\footnote{Since these solutions include different $M_\b/M_\a$, a radial velocity follow-up is also useful to confirm our solution independently of the possible systematics discussed in Section \ref{ssec:phasecurve}.}
Those in panels (c) and (d) yield too short durations for the third eclipse due to the 
head-on crossing with one of the inner binary.
Moreover, the solutions are unphysical because the values of $a_\out/R_\a$ are so small 
that $M_\c/M_\a<0$ is required in Equation (\ref{eq:ac/ra}).
The solution in panel (b), which is the retrograde version of the best solution,  
fits the light curve better than those in (c) and (d);
however, large residuals remain around the first and third tertiary eclipses
because $R_\b$ is slightly smaller than $R_\a$.

Similarly to $F_\b/F_\a$, the constant $A_0$ could also be related to the third-body temperature by 
$T_\c/T_\a=A_0^{1/4}(R_\c/R_\a)^{-1/2}$, which is also listed in Table \ref{tab:results}. 
The value of $T_\c/T_\a$ thus determined, however, should be considered as a rough upper limit
because $A_0$ includes contaminations from nearby sources
and/or systematics in the phase-curve modulation. 

\subsection{Absolute Dimensions}
Combined with the ETV amplitude in Equation (\ref{eq:Aetv}), the mass ratios above
can be further used to determine the absolute masses of the system as
\begin{equation}M_\a=1.074\times10^{-3}M_\odot\left(\frac{A_\mathrm{ETV}}{\mathrm{s}}\right)^3\left(\frac{P_\out}{\mathrm{day}}\right)^{-2}\frac{(1+M_\b/M_\a+M_\c/M_\a)^{2}}{(M_\c/M_\a)^3\sin^3i_\out}.\label{eq:m_abs}\end{equation}
Correspondingly, absolute radii are obtained from 
$a_\mathrm{in}=[P_\mathrm{in}^2GM_\a(1+M_\b/M_\a)/4\pi^2]^{1/3}$ and $a_\mathrm{in}/R_\a$.
The constraints on the absolute dimensions, however, are very weak (see Table \ref{tab:results})
due to the strong correlation $M_\a\sim(M_\c/M_\a)^{-3}\sim(a_\out/R_\a)^{-9}$
as implied by Equations (\ref{eq:mc/ma}) and (\ref{eq:m_abs}).

The constraints are significantly improved
with a better constraint on either $M_\a$ or $M_\c/M_\a$.
To demonstrate this, we repeat the above joint analysis
with the Gaussian prior on the primary mass $M_\a=1.15\pm0.28{\secq}M_\odot$ 
based on the value in the {\it Kepler} Input Catalog (KIC).
Here $M_\a$ and $M_\c/M_\a$ are chosen to be fitting parameters instead of $a_\out/R_\a$ and $A_\mathrm{ETV}$,
where the former two are converted to the latter using Equations (\ref{eq:Aetv}) and (\ref{eq:ac/ra}).
The results are summarized in the last column of Table \ref{tab:results}.
While the constraints on the geometry and relative dimensions are almost unchanged, 
the absolute masses and radii of all three stars are now determined to the precision similar to the prior constraint.
If we also adopt the KIC effective temperature for the primary,
we obtain $T_\mathrm{A}=T_\mathrm{B}=6100\pm200{\secq}\mathrm{K}$ and $T_\mathrm{C}<5000{\secq}\mathrm{K}$.
The dimensions are consistent with the Dartmouth isochrone \citep{2008ApJS..178...89D} of $\sim 7\mathchar`-8{\secq}\mathrm{Gyr}$
and suggest that the inner two stars have entered the subgiant branch and that the third body is an M dwarf \citep{2013AJ....145..102L},
though the conclusion is sensitive to the priors on $M_\a$ and $T_\a$.

\section{Discussion}\label{sec:discussion}
In this Letter, we determine the geometry and physical properties of the hierarchical triple system KIC 6543674
using the {\it Kepler} photometry alone.
Especially, the tertiary event analyzed here enables us to obtain
(i) mutual inclination between the inner and outer binary planes, and
(ii) mass ratio of the inner binary and instantaneous orbital velocity of the third body.
Our analysis clarifies the value of the tertiary eclipses in hierarchical systems
with the clear and textbook-like example of the event.
The methodology presented here is basically applicable to other hierarchical systems 
involving tertiary eclipses on both of the inner stars,
though more sophisticated models of the eclipse light curve and/or ETVs may be required
to accurately model those systems with smaller $P_\mathrm{in}$ and/or $P_\mathrm{out}/P_\mathrm{in}$.
Here it is worth noting that the KIC 6543674 system has the longest
$P_\mathrm{out}$ among the known triply eclipsing hierarchical triples.

The flatness of the system we find (within a few degrees) may have interesting implications
for the  the origin of the closest binaries,
though it is not clear at this point how it compares to the large sample of
misaligned triples \citep{2013ApJ...768...33R, 2015MNRAS.448..946B}
as predicted by the KCTF scenario.
In this context, a large eccentricity of the third body is intriguing
because it may argue for the excitation of the inner orbit's eccentricity on the octupole order
\citep[e.g.,][]{2014ApJ...785..116L, 2015ApJ...805...75P}.
In any case, the relative/absolute dimensions of the system as constrained here
will be useful for testing those possible alternatives.

Although the absolute dimensions derived above are based on the KIC value, which is of limited reliability,
they can be made more accurate with the follow-up spectroscopy 
to better constrain the stellar photospheric parameters and/or to measure radial velocities, 
even if they only cover the inner binary orbit.
In addition, follow-up photometry of another tertiary event will pin down $P_\mathrm{out}$
far more precisely, and can also give us some insight into the dynamical interaction in the system.
In fact, the non-detection of the second tertiary event in the {\it Kepler} data,
which would have occurred around $\mathrm{BJD}=2456114\pm5$ from our result,
suggests that the actual period is $\sim 2\sigma$ longer than our estimate
and that the second event was hidden in the data gap of about $6{\secq}\mathrm{days}$ 
centered around $\mathrm{BJD}=2456126$.
The fact also motivates the ground-based observation of the next event,
which would be around 
July in 2015.
\acknowledgements
We are grateful to the entire {\it Kepler} team for their revolutionary data;
Takayuki Kotani and Shin'ya Yamada for fruitful discussions; and an anonymous referee for many helpful suggestions.
This work is supported by JSPS Grant-in-Aid for Scientific Research No.\secq26-7182 (K.M.) and 25800106 (H.K.) 
and by Research Center for the Early Universe (RESCEU).
K.M. is also supported by the Leading Graduate Course for Frontiers of Mathematical Sciences and Physics.

\bibliographystyle{apj}


\end{document}